\documentclass[10pt]{iopart}
\makeatletter
\@namedef{ver@amsmath.sty}{}
\makeatother
\usepackage{amstext}
\usepackage{graphics}
\usepackage{graphicx}
\usepackage{cite}
\usepackage{epsfig}
\usepackage{epstopdf}
\usepackage{subfigure}
\usepackage{braket}
\DeclareMathAlphabet{\mathpzc}{OT1}{pzc}{m}{it}
\usepackage[dvips]{color}
\usepackage{gensymb}
\usepackage{dcolumn}
\usepackage{bm}
\usepackage{amsmath}
\usepackage{amssymb}
\usepackage{indentfirst}
\usepackage{booktabs}
\usepackage{multirow}
\usepackage{colortbl}

\begin{document}

\title{Magnetism and Magneto-optical Effects in Bulk and Few-layer CrI$_3$: A Theoretical GGA + U Study}

\author{Vijay Kumar Gudelli$^1$ and Guang-Yu Guo$^{2,1}$}
\address{$^1$ Physics Division, National Center for Theoretical Sciences, Hsinchu 30013, Taiwan}
\address{$^2$ Department of Physics and Center for Theoretical Physics, National Taiwan University, Taipei 10617, Taiwan}

\ead{gyguo@phys.ntu.edu.tw}

\vspace{10pt}
\date{\today}

\begin{abstract}
The latest discovery of ferromagnetism in atomically thin films of semiconductors
Cr$_2$Ge$_2$Te$_6$ and CrI$_3$ has unleashed numerous opportunities
for fundamental physics of magnetism in two-dimensional (2D) limit and also 
for technological applications based on 2D magnetic materials.
To exploit these 2D magnetic materials, however, the mechanisms that control
their physical properties should be thoroughly understood.
In this paper, we present a comprehensive theoretical study of the magnetic, 
electronic, optical and magneto-optical properties
of multilayers [monolayer (ML), bilayer and trilayer] as well as bulk CrI$_3$, based on the density functional theory
with the generalized gradient approximation plus on-site Coulomb repulsion scheme.
Interestingly, all the structures are found to be single-spin ferromagnetic semiconductors.
They all have a large out-of-plane magnetic anisotropy energy (MAE) of $\sim$0.5 meV/Cr,
in contrast to the significantly thickness-dependent MAE in multilayers of Cr$_2$Ge$_2$Te$_6$.
These large MAEs suppress transverse spin fluctuations 
and thus stabilize long-range magnetic orders at finite temperatures down to the ML limit.
They also exhibit strong magneto-optical (MO) effects with their Kerr and Faraday rotation 
angles being comparable to that of best-known bulk MO materials. 
The shape and position of the main features in the optical and MO spectra
are found to be nearly thickness-independent although the magnitude of Ker rotation angles increases
monotonically with the film thickness.
Magnetic transition temperatures estimated based on calculated exchange coupling parameters,
calculated optical conductivity spectra, MO Kerr and Faraday rotation angles agree quite well
with available experimental data. The calculated MAE as well as optical and magneto-optical properties
are analyzed in terms of the calculated orbital-decomposed densities of states, band state symmetries and
dipole selection rules.
Our findings of large out-of-plane MAEs and strong MO effects in these single-spin ferromagnetic 
semiconducting CrI$_3$ ultrathin films suggest that they will find valuable applications in 
semiconductor MO and spintronic nanodevices. 
\end{abstract}

%
\ioptwocol

\section{Introduction}
The recent discovery of intrinsic ferromagnetism in atomically thin films of semiconductors 
Cr$_2$Ge$_2$Te$_6$~\cite{gong2017} and CrI$_3$~\cite{huang2017} has opened numerous exciting opportunities 
in two-dimensional (2D) magnetism. Amalgamation of magnetism and 2D materials are highly desirable for at least two reasons, 
one for exploration of fundamental physics of magnetism in 2D limit and the other 
for fascinating technological applications ranging from magnetic memories, to sensing, to spintronics 
to magneto-optical devices based on 2D materials. Therefore, these atomically thin magnetic materials are
currently subject to intensive investigations.
In fact, bulk CrI$_3$ has already been well studied experimentally in the past~\cite{dillon1965,dillon1966,suits_ieee} 
mainly because it is a layered ferromagnetic semiconductor with large Kerr and Faraday rotations
which promise such device applications as optical insulators, magnetic sensors and rewritable optical 
memories~\cite{Dillon1990}. 
Its 2D magnetic structures are presently attracting much renewed interest. 
For example, Seyler {\it et al.}~\cite{seyler-nat-2018} recently reported their observation of magneto-photoluminescence 
in monolayer and bilayer CrI$_3$. Excitingly, manipulation of 2D magnetism 
(e.g., switching of magnetization direction and tuning antiferromagnetic to ferromagnetic
transition) in BL  CrI$_3$ by either applying
a vertical electric field~\cite{Jiang2018a,Huang2018} or electrostatic doping~\cite{Jiang2018b} 
has also been recently demonstrated.
Further, giant tunneling magnetoresistance has been observed in magnetic tunnel junctions made of atomically thin CrI$_3$
and othe van der Waals (vdW) materials.~\cite{Song2018,Wang2018}
Unprecedented control of spin and valley pseudospin in ultrathin CrI$_3$ and WSe$_2$ hetrostructures
has also been demonstrated~\cite{zhong2017}.

In order to exploit these emergent phenomena for various applications, the mechanisms that control
the physical properties of these 2D materials should be thoroughly understood.
As a step towards this goal, we present in this paper a comprehensive first-principles
density functional study of the magnetic, electronic, optical and magneto-optical properties
of multilayers [monolayer (ML), bilayer (BL) and trilayer (TL)] as well as bulk CrI$_3$.
Specifically, we focus the magnetic interactions, magnetic anisotropy energy and magneto-optical
effects in these magnetic structures.
Although magnetic exchange coupling and magnetic anisotropy in ML CrI$_3$ have been theoretically
investigated by several groups based on density functional calculations~\cite{wei2015,Wang2016,Torelli2019,lado2017}, they have
not been studied in BL, TL and bulk CrI$_3$. Furthermore, the interlayer exchange coupling
in bulk and TL CrI$_3$ has not been addressed although that in the BL is presently under intensive
investigations~\cite{Sivadas2018,Jiang-1806,Soriano-1807,Jang-1809}.


Magnetic anisotropy energy (MAE) refers to the energy required to flip the magnetization 
from the easy to the hard axis, and is one of the principal specification parameters for a 
magnetic material. In fact, MAE is particularly important for 2D magnetic materials.
According to the Mermin-Wagner theorem~\cite{mermin1966}, a long-range magnetic order could not occur 
at any finite temperature in a 2D isotropic Heisenberg magnetic structure
because of large thermal spin fluctuations in the 2D system.
However, this Mermin-Wangner restriction~\cite{mermin1966} can be lifted if the 2D system has
a significant out-of-plane MAE which would suppress the thermal fluctuations and thus 
stabilize the long-range magnetic order at finite temperature even in the ML limit, 
as was observed recently in ML CrI$_3$~\cite{huang2017}. The MAE of a magnetic solid arises 
from two contributions, namely, the magnetocrystalline anisotropy energy (MCE) ($\Delta E_{b}$) 
due to the effect of electron relativistic spin-orbit coupling (SOC) on the 
band structure, and the magnetic dipolar anisotropy energy (MDE) ($\Delta E_{d}$) due 
to the magnetostatic dipole-dipole interaction in the magnetic solid. 
In a layered material, the MDE always prefers an in-plane magnetization while 
the MCE could favor either an in-plane or the out-of-plane magnetization depending on
the band structure of the material. Although the MCE of ML CrI$_3$ has been 
calculated~\cite{wei2015,lado2017,Jiang_nl2018}, the MDE was not considered in these previous reports.
Although the MDE is negligibly small for the cubic and isotropic materials such as bcc Fe and fcc Ni, 
it could become significant in low-dimensional materials~\cite{guo91b,tung07,gyg_prb2018}. 
In fact, it was found recently that although the MCE in ML Cr$_2$Ge$_2$Te$_6$ favors the out-of-plane
magnetization, the MDE is larger than the MCE such that ML Cr$_2$Ge$_2$Te$_6$ would have
an in-plane magnetization~\cite{gyg_prb2018}, thereby explaining why the longe-range ferromagnetic order
was not observed in the ML~\cite{gong2017}.
Furthermore, there are no reports on the MAE of bulk, BL and TL CrI$_3$. 
Therefore, in this paper we present both calculated MCE and MDE for 
all the considered CrI$_3$ structures.

\begin{figure}[htb]
\begin{center}
\includegraphics[width=8.5cm]{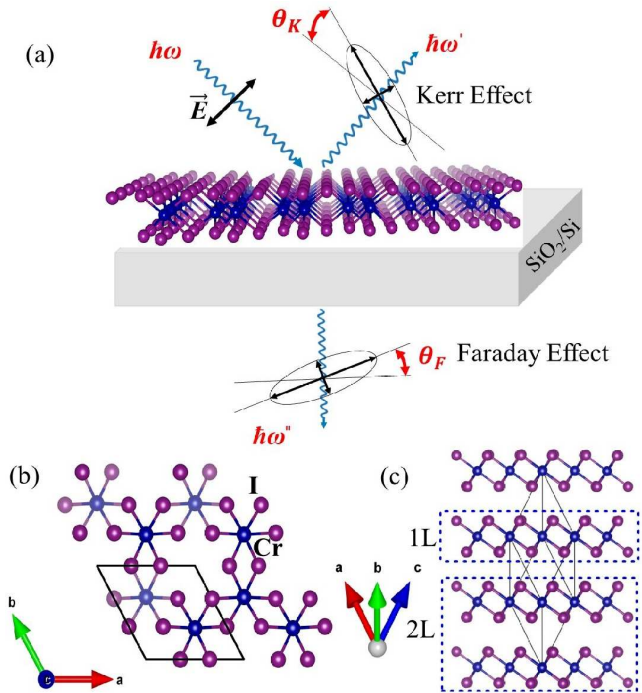}
\caption{(a) Schematic illustration of magneto-optical Kerr and Faraday effects in CrI$_3$. 
(b) ML CrI$_3$ top-view and (c) Crystalline structure of TL CrI$_3$ along with the primitive 
bulk structure (solid black lines) and the blue dashed rectangle indicates ML and BL CrI$_3$ (side-view). 
}
\end{center}
\end{figure}

Magneto-optical (MO) effects are prominent manifestations of light-matter interactions in magnetic materials.
When a linearly polarized light beam hits a magnetic material, 
the reflected and transmitted light beams would become elliptically polarized
with the principal axis rotated with respect to the polarization direction of the incident light beam,
as illustrated in Fig. 1(a). The former and latter are called MO Kerr (MOKE) and MO Faraday (MOFE) 
effects~\cite{antonov2004}, respectively.
MOKE has been widely used to probe the magnetic and electronic properties of solids, 
surface and thin films~\cite{antonov2004}. Indeed, long-range ferromagnetic orders in
atomically thin films of Cr$_2$Ge$_2$Te$_6$ and CrI$_3$ were discovered by using 
the MOKE technique.~\cite{gong2017,huang2017} In contrast, MOFE has been less explored 
mainly because light can only transmit through very thin films. Magnetic materials with large 
Kerr and Faraday rotations are the promising candidates for valuable magneto-optical device 
applications~\cite{mansuripur95,castera96}, and thus they have continuously attracted 
attention in the past several decades. The latest discovery of 2D ferromagnetic semiconductors~\cite{gong2017,huang2017} 
provides especially exciting possibilities of scaling the optical and magneto-optical
devices down to subnanometer scale.
Therefore, here we carry out a systematic first-principles density functional study of the optical
and MO properties of bulk and multilayer CrI$_3$. Indeed, our study reveals that
bulk and multilayer CrI$_3$ exhibit large MO effects in a wide optical frequency range
with Kerr rotation angles being as large as $\sim$2.3$\degree$ and Faraday rotation
angles being in the order of $\sim$200{\degree/$\mu$m}.
This suggests that 2D ferromagnetic CrI$_3$ semiconductor structures will provide 
an interesting material platform for further studies of novel
magneto-optical phenomena and technological applications.

\section{Computational methods}
In this paper, we study the electronic, magnetic, and magneto-optical properties of bulk 
and multilayer CrI$_3$ structures. Bulk CrI$_3$ forms a layered structure with MLs separated 
by the van der Waals gap [Fig. 1(c)]. Each CrI$_3$ ML consists of edge-sharing CrI$_6$ octahedra
forming a planar network with Cr atoms in a honeycomb lattice [Figs 1(b)] and thus
there are two formula units (f.u.) per lateral unit cell. 
These MLs are then stacked in an ABC sequence, 
thus resulting in a rhombohedral crystal with $R\bar{3}$ symmetry and two f.u. 
per unit cell.  This structure can also be regarded as an ABC-stacked hexagonal crystal 
with experimental lattice constants $a=b=6.867$ \AA$ $ and $c=19.807$ \AA~\cite{mag2015}. 
We consider the experimental rhombohedral primitive cell in the bulk calculations. 
For the multilayer CrI$_3$ structures, the hexagonal unit cell [Figs 1(b)] with the experimental 
lattice constants is considered as the lateral unit cell. For BL and TL CrI$_3$ structures 
[see Fig. 1(c)], we consider the AB and ABC stackings, respectively.
The slab-superlattice approach is used to construct the multilayer structures with the 
separations of neighboring slabs being at least 20 \AA.

First-principles density functional calculations are performed using the accurate projector 
augmented wave (PAW) method~\cite{kresse1999ultrasoft} as implemented in the Vienna ab-initio simulation 
package (VASP)~\cite{kresse1996efficient,kresse1996efficiency}. 
The fully relativistic PAW potentials are adopted in order to include the SOC. 
The valence configurations of Cr and I atoms adopted
in the calculations are 3d$^{5}$ 4s$^1$, 5s$^2$ 5p$^5$, respectively.
A large plane-wave energy cutoff of 400 eV is used throughout the calculations.
For the Brillouin zone integrations, $k$-point meshes of 16$\times$16$\times$16 and 20$\times$20$\times$1
are used for bulk and multilayers CrI$_3$, respectively.

The generalized gradient approximation (GGA) to the exchange-correlation potential  
of Perdew-Burke-Ernzerhof parametrization~\cite{pbe_1996} is used. 
However, an improved account of onsite electron correlation among $3d$ electrons
is needed for a better description of the electronic and magnetic structures
of $3d$ transition metal compounds (see, e.g., Ref. \cite{Jeng04} and references therein).
To better describe onsite Coulomb repulsion between Cr $3d$ electrons, we adopt 
the GGA+$U$ scheme~\cite{dudarev98} in the present calculations. 
Feng {\it et al.}~\cite{gyg_prb2018} recently carried out systematic GGA$+U$ calculations for bulk and multilayer 
Cr$_2$Ge$_2$Te$_6$ with different effective Coulomb energy $U$ values and concluded
that the appropriate $U$ value for Cr $3d$ electrons is 1.0 eV.
Given the similarity between the Cr$_2$Ge$_2$Te$_6$ and CrI$_3$, here we also use $U = 1.0$ eV.
Nevertheless, we also perform a series of the GGA$+U$ calculations with the $U$ value ranging
from 0 to 3 eV for all the considered CrI$_3$ systems and find that the calculated physical quantities
do not depend significantly on the $U$ value used [see Supplementary Note A, table S1 and Fig. S1
in the Supplementary Information (SI)]. This indicates that the results presented below 
would remain nearly the same even if a different $U$ value were used. 

We consider four intra-layer magnetic configurations, namely, one FM
as well as three antiferromagnetic (AF) structures, labelled as AF-N$\rm \acute{e}$el, AF-zigzag,
and AF-stripe in Fig. 2. For bulk, BL and TL CrI$_3$, we also consider interlayer FM and AF
magnetic configurations with intralayer FM CrI$_3$ MLs.
To understand the magnetic interactions and also to estimate magnetic ordering temperature 
({\it {T$_c$}}) for bulk and multilayers CrI$_3$, we calculate the exchange coupling parameters 
by mapping the calculated total energies of these magnetic configurations (see Fig. 2) 
onto the classical Heisenberg Hamiltonian, 
$
{\it E=E_0-\sum_{i,j}J_{ij}\hat{\mathbf{e}}_i\ {\bm\cdot}\ \hat{\mathbf{e}}_j},
$
where $E_0$ donates the nonmagnetic ground state energy; $J_{ij}$ the exchange coupling parameter 
between sites $i$ and $j$; $\hat{\mathbf{e}}_i$, the unit vector representing the direction 
of the magnetic moment on site $\it i$. For one CrI$_3$, we obtain a set of four
linear equations of $J_1$, $J_2$ and $J_3$: $E_{FM} = E_0-3J_1-6J_2-3J_3, 
E_{AF-N\rm\acute{e}el} = E_0+3J_1-6J_2+3J_3, E_{AF-zigzag} = E_0-J_1+2J_2+3J_3$ 
and $E_{AF-stripe} = E_0+J_1+2J_2-3J_3$, which can be solved to estimate $J_1$, $J_2$ and $J_3$. 
Similarly, one can estimate the effective interlayer coupling parameter $J_{z}$ based the calculated 
total energies per unit cell of two magnetic configurations by using expressions $J_z = (E_{AF}-E_{FM})/4$ 
for the BL, $J_z = (E_{AF}-E_{FM})/8$ for the TL and $J_z = (E_{FiM}-E_{FM})/8$ for the bulk
where $E_{FiM}$ denotes the total energy of the ferrimagnetic configuration with the Cr magnetic moments
on one of the three layers in the hexagonal unit cell fliped.

We also calculate the MCE, MDE and hence MAE for all the considered structures in the FM state. 
We first perform two relativistic electronic structure calculations for the in-plane and out-of-plane 
magnetizations, and then obtain the MCE ($\Delta E_b$) as the total energy difference between the two calculations. 
Highly dense k-point meshes of 24$\times$24$\times$24 and 28$\times$28$\times$1 are used for bulk 
and multilayers CrI$_3$, respectively. Further calculations by using different k-point meshes 
show that thus-obtained MCEs converge within 1 \%. For a FM system, the MDE $\Delta E_d$ 
in atomic Rydberg units is given by ~\cite{guo91b,tung07}
\begin{equation}
E_d = \sum_{qq'} \frac{2m_q m_{q'}}{c^2}M_{qq'}
\end{equation}
where the speed of light $c=274.072$ and the so-called magnetic dipolar Madelung constant
\begin{equation}
M_{qq'} = \sum_{{\bf R}}^{'}\frac{1}{|{\bf R}+{\bf q}+{\bf q}'|^3}\left\{1-3\frac{[({\bf R}+{\bf q}+{\bf q}').\hat{m}_q]^2}{|{\bf R}+{\bf q}+{\bf q}'|^2}\right\}
\end{equation}
where {\bf R} are the lattice vectors, {\bf q} are the atomic position vectors in the unit cell,
and {\bf $m_q$} is the atomic magnetic moment (in units of $\mu_B$) on site {\bf q}. 
Using the calculated magnetic moments, the $\Delta E_d$ is obtained as the difference in $E_d$ 
between the in-plane and out-of-plane magnetizations. 
In a 2D material, since all the {\bf R} and {\bf q} are in-plane, the second term in Eq. (2) 
would be zero for the out-of-plane magnetization, thereby resulting in the positive $M_{qq'}$ while 
for an in-plane magnetization the $M_{qq'}$ is negative.~\cite{gyg_prb2018} 
Therefore, the MDE always prefers an in-plane magnetization in a 2D material. 
This is purely a geometric effect and consequently the MDE is also known as the magnetic shape anisotropy energy.

In a FM solid with trigonal symmetry and magnetization rotational along $z$-axis, 
the optical conductivity tensor consists of three independent elements, 
namely, $\sigma_{xx}$, $\sigma_{zz}$ and $\sigma_{xy}$. Within the Kubo linear response 
theory~\cite{wang1974band,feng2015large}, the adsorptive parts of these elements
are given by 
\begin{equation}
\sigma_{aa}^{1} (\omega) = \frac{\pi e^2}{\hbar\omega m^2}
\sum_{i,j}\int_{BZ}\frac{d{\bf k}}{(2\pi)^3}|p_{ij}^{a}|^{2}
\delta(\epsilon_{{\bf k}j}-\epsilon_{{\bf k}i}-\hbar\omega),
\end{equation}
\begin{equation}
\sigma_{xy}^{2} (\omega) = \frac{\pi e^2}{\hbar\omega m^2}
\sum_{i,j}\int_{BZ}\frac{d{\bf k}}{(2\pi)^3}\text{Im}[p_{ij}^{x}p_{ji}^{y}]
\delta(\epsilon_{{\bf k}j}-\epsilon_{{\bf k}i}-\hbar\omega),
\end{equation}
where $\hbar \omega$ is the photon energy, and $\epsilon_{{\bf k}i}$ is the $i$th band energy at point ${\bf k}$. 
Summations $i$ and $j$ are over the occupied and unoccupied bands, respectively. Dipole matrix 
elements $p_{ij}^{a} = \langle\textbf{k}\emph{j}|\hat{p}_{a}|\textbf{k}i\rangle$ where $\hat{p}_a$ denotes 
Cartesian component $a$ of the dipole operator, are obtained from the band structures within the PAW formalism~\cite{Adolph01}, 
as implemented in the VASP package. The integration over the Brillouin zone is carried out by using 
the linear tetrahedron method (see Ref. \cite{Temmerman89} and references therein). The dispersive parts 
of the optical conductivity elements can be obtained using the Kramers-Kroing relations
\begin{equation}
\sigma_{aa}^{2} (\omega) = -\frac{2\omega}{\pi }P \int _{0}^{\infty }\frac{\sigma_{aa}^{1}(\omega ')}{\omega^{'2}-\omega ^{2}}d\omega ^{'},
\end{equation}
\begin{equation}
\sigma_{xy}^{1} (\omega) = \frac{2}{\pi }P \int _{0}^{\infty }\frac{\omega^{'}\sigma_{xy}^{2}(\omega ')}{\omega^{'2}-\omega ^{2}}d\omega ^{'}
\end{equation}
where $P$ donates the principle value. 
The quasiparticle lifetime effects are accounted for by convoluting all the optical conductivity
spectra with a Lorentzian of line width $\Gamma$.
For layered van der Waals materials such as graphite, $\Gamma$ is about 0.2 eV (see, e.g.,
Figs. 1(a) and 1(b) in Ref. ~\cite{Guo04}), which is thus used in this paper.

For a bulk magnetic material, 
the complex polar Kerr rotation angle is given by~\cite{gyg_prb50_1994,gyg_prb51_1995},
\begin{equation}
\theta _{K}+i\epsilon _{K}=\frac{-\sigma _{xy}}{\sigma _{xx}\sqrt{1+i(4\pi/\omega)\sigma _{xx}}}.
\end{equation}
However, for a magnetic thin film on a nonmagnetic substrate, the complex polar Kerr rotation 
angle is given by~\cite{suzuki_1992,feng2016}
\begin{equation}
\theta _{K}+i\epsilon _{K}=i\frac{2\omega d}{c}\frac{\sigma _{xy}}{\sigma_{xx}^{s}}=\frac{8\pi d}{c}\frac{\sigma_{xy}}{(1-\varepsilon^s_{xx})}
\end{equation}
where $d$ stands for the thickness of the magnetic layer and $\varepsilon^s_{xx} (\sigma_{xx}^s)$ 
the diagonal part of the dielectric constant (optical conductivity) of the substrate. Experimentally, 
atomically thin CrI$_3$ films were prepared on SiO$_2$/Si~\cite{huang2017,zhong2017,seyler-nat-2018}. 
Thus, the dielectric constant ($\varepsilon^s_{xx} = 2.25$) of bulk SiO$_2$ is used here.

Similarly, the complex Faraday rotation angle for a thin film can be written as~\cite{ravindran_1999}
\begin{equation}
\theta _{F}+i\epsilon _{F}=\frac{\omega d}{2c}(n_{+}-n_{-})
\end{equation}
where $n_+$ and $n_-$ represent the refractive indices for left- and right-handed polarized lights, respectively, 
and are related to the optical conductivity (or dielectric function) via expressions 
$n_{\pm}^{2}=\varepsilon_{\pm}=1+{\frac{4\pi i}{\omega}}\sigma _{\pm}=1+{\frac{4\pi i}{\omega}}(\sigma _{xx}\pm i \sigma _{xy})$.
Here the real parts of the optical conductivity $\sigma_{\pm}$ can be written as 
\begin{equation}
\sigma^1_\pm(\omega) = \frac{\pi e^2}{\hbar\omega m^2}
\sum_{i,j}\int_{BZ}\frac{d{\bf k}}{(2\pi)^3} |\Pi_{ij}^{\pm}| ^{2}
\delta(\epsilon_{{\bf k}j}-\epsilon_{{\bf k}i}-\hbar\omega)
\end{equation}
where $\Pi _{ij}^\pm = \langle\textbf{k}\emph{j}|\frac{1}{\sqrt{2}}(\hat{p}_{x}\pm \emph{i}\hat{p}_{y})|\textbf{k}i\rangle$. 
Clearly, $\sigma _{xy} = \frac{1}{2i}(\sigma _+-\sigma _-)$, and thus $\sigma _{xy}$ would be nonzero 
only if $\sigma _+$ and $\sigma _-$ are different. In other words, magnetic circular dichroism is essential 
for the nonzero $\sigma _{xy}$ and hence the magneto-optical effects.



\begin{figure}[htb]
\begin{center}
\includegraphics[width=8.5cm]{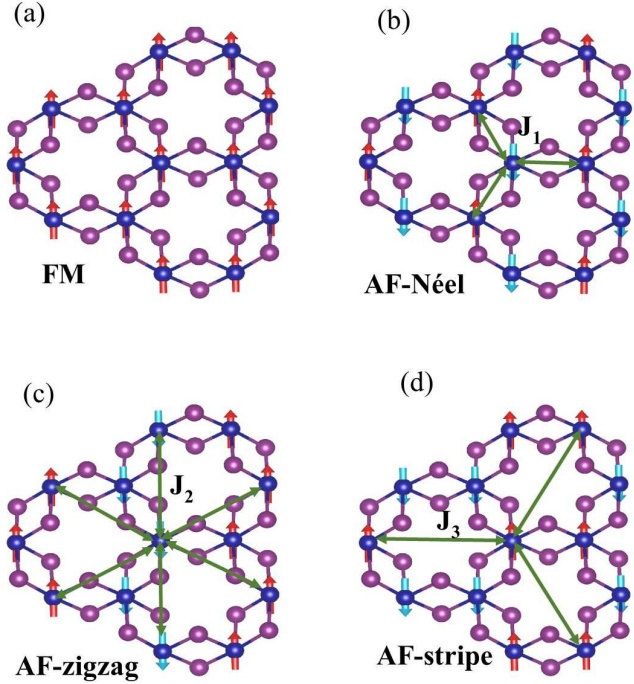}
\end{center}
\caption{Four considered intralayer magnetic configurations: (a) FM, (b) AF-N$\rm\acute{e}$el, 
(c) AF-zigzag, (d) AF-stripe types. The magnetic Cr atoms are shown with red (blue) arrows indicating up (down) spins. 
Intralayer exchange coupling parameters $J_1$, $J_2$, and $J_3$ are indicated by green arrows. 
}	
\end{figure}

\section{Results and discussion}
\subsection{Magnetic properties}
\subsubsection{Magnetic structure and exchange coupling:} 

To find the ground state magnetic structure and also evaluate exchange coupling among the magnetic
Cr atoms in all the considered CrI$_3$ structures, we first consider four possible intra-layer magnetic 
configurations, namely, the FM, N$\rm \acute{e}$el-type antiferromagnetic (AF-N$\rm \acute{e}$el), 
zigzag-antiferromagnetic (AF-zigzag) and stripe-like antiferromagnetic (AF-stripe) structures (see Fig. 2). 
The calculated total energies for these magnetic configurations indicate that 
for all the considered structures, the FM state is the most stable configuration.
We then consider the interlayer ferromagnetic and antiferromagnetic [see Fig. 2(e)] structures for
bulk, BL and TL CrI$_3$, and again we find the interlayer FM state is more stable.
Therefore, we list the calculated spin and orbital magnetic moments of bulk and multilayer CrI$_3$ 
in the FM state only in Table 1. The calculated Cr spin magnetic moment in all the CrI$_3$ structures 
is $\sim$3.2 $\mu_B$, being in good agreement with the experimental value of $\sim$3.0 $\mu_B$~\cite{mag2015}. 
This is also consistent with three unpaired electrons in the Cr $t_{2g}$ configuration in these 
structures. Further, the calculated Cr orbital magnetic moment for all the investigated structures 
is small ($\sim$ 0.08 $\mu_B$), thus lending support to the notion of the completely filled spin-up 
Cr $t_{2g}$ subshell in these systems. Interestingly, there is also a small induced magnetic moment
(-0.09 $\mu_B$) on the I atom (Table 1), which is antipallel to that of the Cr magnetic moment,
in all the systems. All these result in a total magnetic moment of 3.00 $\mu_B$/f.u. for all the structures.
\begin{table*}[htbp]
\begin{center}
\caption{Total spin magnetic moment ($m_s^t$), atomic (averaged) spin ($m_s^{Cr}$, $m_s^{I}$) 
and orbital ($m_o^{Cr}$, $m_o^{I}$) magnetic moments as well as band gap ($E_g$), magnetocrystalline anisotropy
energy ($\Delta E_{b}$), dipolar anisotropy energy ($\Delta E_{d}$) and (total) magnetic anisotrpy
energy ($\Delta E_{ma} = \Delta E_{b} + \Delta E_{d}$) 
of bulk and few-layer CrI$_3$ in the ferromagnetic 
ground state (magnetization being perpendicular to the layers) calculated using the GGA+U scheme 
with the spin-orbit coupling included. Also listed are the experimental magnetic anisotropy 
energy ($\Delta E_{ma}^{exp}$) and band gap ($E_g^{exp}$) for comparison.
}

\begin{tabular}{ccccccccc}
\br 
           & $m_s^t$ & $m_s^{Cr}$ ($m_o^{Cr}$) & $m_s^{I}$ ($m_o^{I}$) & 
$\Delta E_{b}$ ($\Delta E_{d}$) & $\Delta E_{ma}$ ($\Delta E_{ma}^{exp}$) & $E_g$ ($E_g^{exp}$)   \\
     &  ($\mu_B$/f.u.) & ($\mu_B$/atom) & ($\mu_B$/atom) & (meV/f.u.) & (meV/f.u.) & (eV) \\
\hline
  ML & 6.00 & 3.20 (0.083) & -0.092 (-0.014) &   0.678 (-0.152) & 0.526 & 0.81  \\
  BL & 6.00 & 3.21 (0.083) & -0.094 (-0.014) &   0.620 (-0.152) & 0.468 & 0.71  \\
  TL & 6.00 & 3.21 (0.083) & -0.093 (-0.014) &   0.586 (-0.152) & 0.434 & 0.68  \\
Bulk & 6.00 & 3.21 (0.082) & -0.094 (-0.014) &   0.545 (-0.064) & 0.481 (0.26$^a$) & 0.62 (1.2$^a$) \\ 
\br
\end{tabular} \\
$^a$Ref.~\cite{dillon1965}.
\end{center}
\end{table*}

As mentioned in Sec. 2, using the total energies of the considered magnetic configurations,
we estimate first three near-neighbor exchange coupling parameters ($J_1$, $J_2$, $J_3$)
in each ML in all the considred systems and also the effective interlayer exchange 
coupling constant ($J_{z}$). 
The calculated exchange coupling constants are listed in Table 2. 
It is clear from Table 2 that the first near-neighbor exchange coupling ($J_1$) 
is ferromagnetic (i.e., positive) in all the considered CrI$_3$ structures.  
The second near-neighbor coupling ($J_2$) is also ferromagnetic (positive),
although the $J_2$ value is about 3.5 times smaller than $J_1$ for all the structures. 
Interestingly, Table 2 shows that the magnitudes of both $J_1$ and $J_2$ increase 
monotonically with the number of layers,
being consistent with the observed increase in magnetic transition
temperature in these systemsi\cite{huang2017}. 
The third near-neighbor magnetic coupling ($J_3$) is also ferromagnetic (positive)
in  all the CrI$_3$ structures except ML CrI$_3$ where $J_3$ is negative (antiferromagnetic). 
Nonetheless, the calculated $J_3$ values are found to be 5 times smaller than $J_1$ and only 
about half of $J_2$ for all the structures. 
For the CrI$_3$ ML, we find that all the calculated exchange coupling parameters are
in good quantitive agreement with the recent GGA calculations~\cite{wei2015,Wang2016,Torelli2019},
although the $J_1$ is slightly larger than that of the previous GGA+$U$ calculation\cite{lado2017}
with $U=2.7$ eV and $J=0.7$ eV. Note that only the first near-neighbor magnetic interaction
was considered in Ref. \cite{lado2017} and also that the $J$ values reported previously~\cite{wei2015,Wang2016,Torelli2019,lado2017}
should be multiplied by a factor of $S^2 = 9/4$ in order to compare with the present $J$ values.

Finally, we notice that the calculated interlayer exchange coupling constant $J_{z}$ is also ferromagnetic 
for bulk, BL and TL structures. This prediction agrees well with all previous 
experiments~\cite{huang2017,dillon1965,dillon1966,suits_ieee} on all the structures except BL CrI$_3$
which has been experimentally found to be interlayer antiferromagnetically coupled~\cite{huang2017}. 
A microscopic explanation of this notable discrepancy between theory and experiment on BL CrI$_3$
is nontrivial. Indeed, very recently this problem has already been investigated theoretically 
by several groups~\cite{Sivadas2018,Jiang-1806,Soriano-1807,Jang-1809}.
These theoretical studies all showed that the interlayer magnetic coupling in BL CrI$_3$
is strongly stacking dependent and is also sensitive to the exchange-correlation functional used. 
In particular, the GGA + $U$ calculation with $U = 3$ eV~\cite{Sivadas2018}
indicates that in the AB$'$-stacked BL CrI$_3$, the total energy of the AF configuration is 
slightly lower than that of the FM configuration (by merely $\sim$0.7 meV/unit cell). 
However, the AB-stacking order which is the global minimum and favors the FM state, 
still have a total energy significantly lower than the AB$'$-stacked structure (by $\sim$6 meV/unit cell).
Consequently, it was proposed~\cite{Sivadas2018} that BL CrI$_3$ exfoliated at room-temperature
is perhaps kinetically trapped in the metastable AB$'$-stacked structure, on cooling 
and thus the AF state was observed instead~\cite{huang2017}. Furthermore, it was also found that
even in the AB$'$-stacking order, the AF interlayer coupling becomes favorable only
when $U$ is larger than 2.5 eV,~\cite{Jiang-1806} and this result is confirmed by our own
GGA + $U$ calculations. Clearly, further measurements
on the stacking order in BL CrI$_3$ are needed to resolve this interesting problem.
To this end, we calculate the optical and magneto-optical properties of BL CrI$_3$ in both
AB- and AB$'$-stacking orders. Unfortunately,
we find that the magneto-optical effects of BL CrI$_3$ hardly depend on the stacking order,
as presented in Figs. S4 and S5 in the supplementary information.
\begin{table*}
\begin{center}
\caption{
Intralayer first-, second- and third-neighbor exchange coupling parameters ($J_1$, $J_2$, $J_3$) 
as well as effective interlayer exchange coupling constant ($J_{z}$) in ML, BL, TL 
and bulk CrI$_3$, derived from the calculated total energies for various magnetic configurations 
(see the text). Ferromagnetic transition temperatures estimated using the derived $J$ values 
within the mean-field approximation ($T_c^m$) and also within the spin wave theory (SWT) 
($T_c^{SWT}$) (see the text) are also included. 
For comparison, the available experimental $T_c$ values are also listed in brackets.}
\begin{tabular}{ccccccc}
\br
          &  $J_1$ &$J_2$ & $J_3$ &$J_{z}$ & $T_c^m$ & $T_c^{SWT}$ ($T_c^{exp}$) \\
          & (meV) &(meV) & (meV) &  (meV) & (K) & (K) \\ \hline
 ML 	  &6.91 & 1.48 & -0.46 & -    & 109&  66 (45$^a$) \\
 BL       &7.59 & 2.30 &  1.02 & 2.97 & 165&  69 \\
 TL       &8.05 & 2.45 &  0.91 & 2.85 & 172&  71 \\
Bulk      &8.08 & 2.89 &  1.62 & 2.95 & 191&  73 (61$^b$) \\
\br
\end{tabular}\\
$^a$Ref.~\cite{huang2017}, $^b$Ref.~\cite{wei2015}.
\end{center}
\end{table*}

\subsubsection{Magnetic anisotropy energy and ferromagnetic transition temperature:}
The calculated magnetic anisotropy energies ($\triangle E_{ma}$) for bulk and multilayers CrI$_3$
are presented in Table 1. As mentioned before, the MAE consists of two competing contributions,
namely, magnetocrystalline anisotropy energy ($\triangle E_{b}$) due to the SOC effect
on the band structure and magnetic dipolar (shape) anisotropy energy ($\triangle E_{d}$). 
First of all, Table 1 shows that as could be expected from the 2D nature of
their structures, the MDE per ML in all the considered systems is rather large 
and independent of the film thickness. Furthermore, the MDE is always negative
and thus always prefers an in-plane magnetization.
Remarkably, the MCE per ML is a few times larger than the MDE, and positive, 
meaning that it prefers the out-of-plane magnetization. The sum of these competing terms
($\triangle E_{b} +\triangle E_{d}$) thus results in the large and positive MAE
(Table 1), which is more or less independent of the film thickness. 
This is in strong contrast to the case of the Cr$_2$Ge$_2$Te$_6$ multilayers 
where the MAE is much smaller and depends significantly on the film thickness~\cite{gyg_prb2018}.
In particular, for ML Cr$_2$Ge$_2$Te$_6$, the magnitude of MCE becomes smaller than
that of MDE such that the MAE is negative, favoring an in-plane magnetization~\cite{gyg_prb2018}.
Importantly, since the significant out-of-plane MAE is needed to stabilize long-range magnetic orders
in a 2D material by suppressing the thermal spin fluctuations, this explains why
the FM state is observed in ML CrI$_3$~\cite{huang2017} but not in ML Cr$_2$Ge$_2$Te$_6$~\cite{gong2017}. 
These contrasting magnetic anisotropic properties between
multilayers CrI$_3$ and Cr$_2$Ge$_2$Te$_6$ thus lend support to the notion
that the former is well described by the 2D Ising model while the latter is well
described by the Heisenberg model.\cite{Samarth2017}.
Note that the present MCE of ML CrI$_3$ is in good quantitative agreement with 
the theoretical  value of 0.686 meV/f.u. reported recently in Ref. ~\cite{wei2015} in which the MDE was not calculated. 
Also, the calculated MAE of bulk CrI$_3$ agree reasonably well with the experimental value reported
in Ref. \cite{dillon1965} (see Table 1).
Since the MAEs of the CrI$_3$ multilayers are large and comparable to heavy element magnetic
alloys such as FePt and CoPt ~\cite{Oppeneer98}, which have the largest MAEs among magnetic transition metal
alloys and compounds, these CrI$_3$ multilayers could 
have valuable applications in high density data storage devices.


Now we consider the ferromagnetic transition temperature ($T_c$) of the  CrI$_3$ systems, 
which is another important parameter for a magnetic material. 
A rough estimation of $T_c$ could be made within the mean-field approximation (MFA), 
i.e., $k_BT_c^m = \frac{1}{3}J_0$ where $J_0 = \sum_{j}J_{0j}$ and the present cases $J_0 = 3J_1 + 6J_2 + 3J_3 + J_{z}$.~\cite{Halilov98} 
The estimated ferromagnetic transition temperatures ($T_c^m$) are listed in Table 2. 
It is clear from Table 2 that the estimated $T_c^m$ and measured $T_c$ agree in trend, 
i.e., the magnetic transition temperature increases monotonically from ML, to BL, to TL and to bulk  CrI$_3$.
Nonetheless, the estimated $T_c^m$ values are much higher.  In particular, for ML CrI$_3$, the estimated 
$T_c^m$ of 109 K is more than two times larger than the experimental $T_c$ 45 K for ML CrI$_3$,
and in bulk CrI$_3$, the $T_c^m$ of 191 K is about three times larger than the measured $T_c$ 61 K.\cite{huang2017}.
The reason for this is that the MFA neglects transverse
spin fluctuations~\cite{Pajda00,Irkhin99}, which is especially strong in low-dimensional materials. 
Furthermore, the MFA violates Mermin-Wagner theorem~\cite{mermin1966} which says that long-range magnetic order at
a finite temperature cannot exist in an isotropic 2D Heisenberg magnet.
However, the out-of-plane MAE in 2D materials 
can establish long-range magnetic orders at finite temperatures by opening a gap in 
the spin wave spectrum which would suppress thermal spin fluctuations~\cite{Pajda00,Irkhin99}. 
Therefore, a better estimation of the $T_c$ would be based on the spin wave theory (SWT) with  
the out-of-plane MAE taken into account. This leads to an approximate expression of $T_c$
as $k_BT_c^{SWT} = \pi J_1 / 2 log(1 + 2\pi J_1 / \Delta_0)$ where $\Delta_0$ is the spin wave gap~\cite{lado2017},
which can be roughly set to $\Delta_0 = \Delta E_{ma}$. The $T_c^{SWT}$ values 
evaluated using the calculated $J_1$ and $\Delta E_{ma}$ are listed in Table 2. 
It can be seen that compared with $T_c^{m}$, the $T_c^{SWT}$ values agree better with 
the corresponding experimental values although they are still too high (Table 2).

\subsection{Electronic structure}
Better understanding of the electronic, magnetic and magneto-optical properties of the materials 
can be obtained from the detailed analysis of the electronic band structure. Therefore, fully relativistic 
electronic band structure of bulk and ML CrI$_3$ structures are displayed in Fig. 3, 
while that of the BL and TL are displayed in Fig. S2 in the supplementary information (SI). 
Notably, these figures show that all the multilayer structures are almost direct band-gap semiconductors 
with both valance band maximum (VBM) and conduction band minimum (CBM) located at $\Gamma$ point. 
In contrast, the bulk structure is an indirect bandgap semiconductor with the VBM sitting at the $\Gamma$ point 
and the CBM located at the T point. The obtained semiconducting band gaps for all the structures are listed in Table 1,
which shows that the band gap increases slightly as bulk CrI$_3$ is thinned down to TL, to BL and finally to ML. 

\begin{figure}[htb]
\begin{center}
\includegraphics[width=8cm]{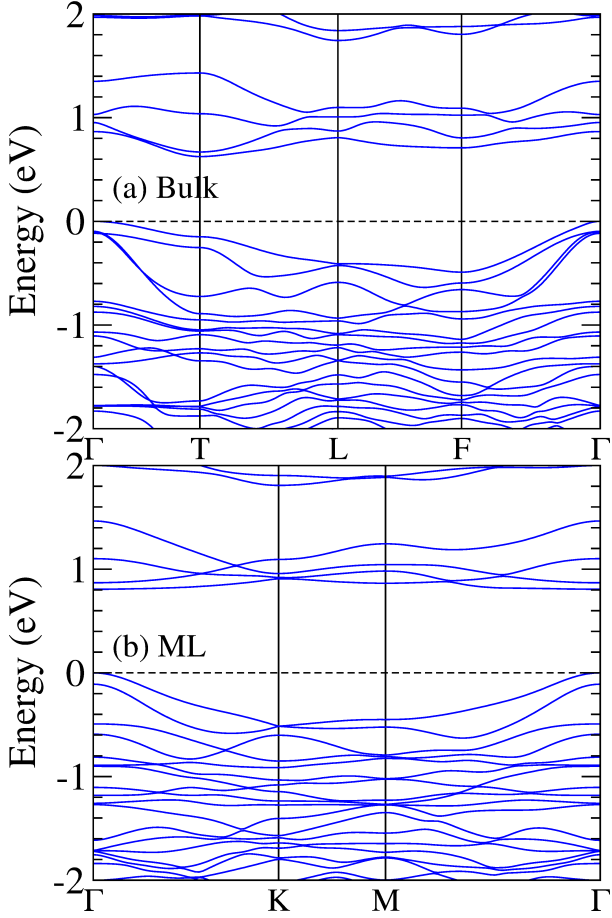}
\end{center}
\caption{Relativistic band structures of (a) bulk and (b) ML CrI$_3$, respectively. 
Horizontal dashed lines denote the top of valance band.}
\end{figure}

\begin{figure}[htb]
\begin{center}
\includegraphics[width=8cm]{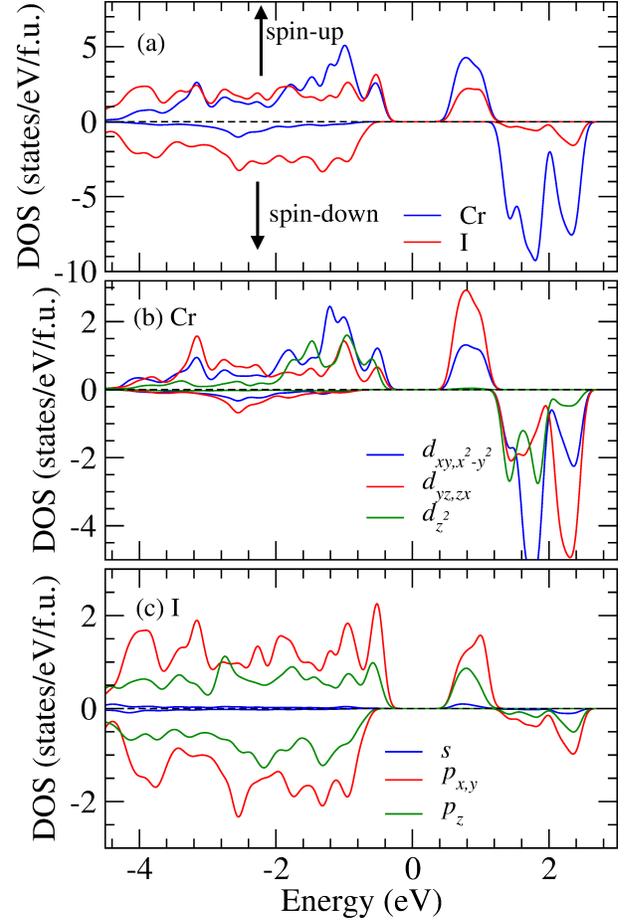}
\caption{Site-, orbital-, and spin-projected scalar-relativistic densities of states (DOS) of bulk CrI$_3$}
\end{center}
\end{figure}

To analyze the spin characters of the bands at the VBM and CBM,
we present the spin-polarized scalar-relativistic band structures of all the structures considered here
in Fig. S3 in the SI. Figure S3 indicates that both the lower conduction bands and top valence bands
of all the structures are of purely spin-up character. Interestingly, this shows that 
all the structures are single-spin semiconductors and thus would be promising candidates
for efficient spintronic and spin photovoltaic devices (see Ref. \cite{Cai2017} and references therein). 
We also calculate the total as well as site, orbital projected spin-up and spin-down densities of states 
(DOS) for all the CrI$_3$ structures. The calculated DOS spectra of bulk and monolayer CrI$_3$ are displayed
in Figs. 4 and 5, respectively, as representatives. Figure 4 shows that the contributions to the valance and conduction bands
in the energy ranges of -4.5 to -0.3 eV and 0.4 eV to 2.5 eV, respectively, come mostly from the Cr $d$ orbitals 
with significant contributions from the I $p$ orbitals as well. The contributions from both the Cr and I confirm 
the strong hybridization between Cr $d$ and I $p$ orbitals in CrI$_3$ structures. The spin-up valance band states 
are predominately arise from the Cr $d$ orbitals, along with minor contribution with the I $p$ orbitals 
at the valence band edge. Conversely, the DOS states in spin-down valance band region are mostly contributed 
from I $p$ orbitals. The states in the upper valance band region of -4.5 eV to -0.3 eV are made up 
of spin-up Cr $d_{xy,x^2-y^2}$ orbitals, with a broad peak at 1.3 eV. In case of conduction bands ranging 
from 0.4 eV to 1.2 eV is mainly because of spin-up Cr $d_{xz,yz}$ orbitals. From these one could
say that the band gap in the CrI$_3$ structures arise due to the crystal-field splitting of 
spin-up Cr $d_{xy,x^2-y^2}$ and $d_{xz,yz}$ bands. The spin-down conduction bands appear only above 1.3 eV
with pronounced peaks made up of mainly Cr $d_{xy,x^2-y^2}$ and $d_{z^2}$ orbitals (Fig. 4). 
Furthermore, the calculated total as well as site, orbital projected spin-up and spin-down DOS 
spectra of ML CrI$_3$ (Fig. 5) are rather similar to that of bulk CrI$_3$ (Fig. 4). 
The main difference comes from the contributions of the I $p$ orbital 
which is enhanced in the higher energy region in the ML. 

\begin{figure}[htb]
\begin{center}
\includegraphics[width=8cm]{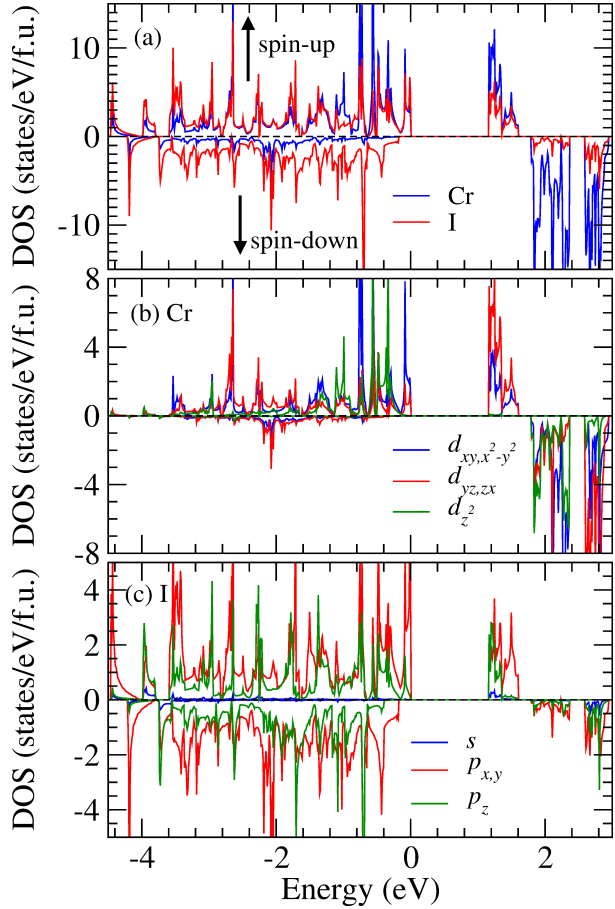}
\caption{Site-, orbital-, and spin-projected scalar-relativistic densities of states (DOS) of ML CrI$_3$.}
\end{center}
\end{figure}

As mentioned above, the insulating band gap in the CrI$_3$ structures arises due to the crystal-field 
splitting of Cr $d_{xy,x^2-y^2}$ and $d_{xz,yz}$ states. The symmetries of these states near the band gap edges 
are important to the magneto-crystalline anisotropy in the presence of SOC~\cite{tung07,wang1993}, 
according to the perturbation theory. In particular, the SOC matrix elements for the $d$ orbital 
of $\langle d_{xz}|H_{SO}|d_{yz} \rangle$ and $\langle d_{x^2-y^2}|H_{SO}|d_{xy} \rangle$ would 
favor the perpendicular magnetic anisotropy, while 
other elements of $\langle d_{x^2-y^2}|H_{SO}|d_{yz} \rangle$, $\langle d_{xy}|H_{SO}|d_{yz} \rangle$
and $\langle d_{z^2}|H_{SO}|d_{yz} \rangle$ would prefer an in-plane magnetic anisotropy~\cite{takayama_1976}. 
The magnitude ratio of these $d$-orbital SOC matrix elements are 
$\langle d_{xz}|H_{SO}|d_{yz} \rangle^2$: $\langle d_{x^2-y^2}|H_{SO}|d_{xy}\rangle^2$: $\langle d_{x^2-y^2}|H_{SO}|d_{yz}\rangle^2$: $\langle d_{xy}|H_{SO}|d_{yz}\rangle^2$: $\langle d_{z^2}|H_{SO}|d_{yz}\rangle^2$ = $1:4:1:1:3$~\cite{takayama_1976}. 
Figures 4 and 5 show that in the upper valence and lower conduction band region,
the $d_{xz,yz}$ and $d_{x^2-y^2,xy}$ orbital-decomposed DOSs are large and this would result in
large $\langle d_{xz}|H_{SO}|d_{yz} \rangle$ and $\langle d_{x^2-y^2}|H_{SO}|d_{xy} \rangle$
and hence large contributions to the perpendicular anisotropy. In contrast, the $d_{z^2}$ DOS is almost zero 
in the lower conduction band region of 0.4$\sim$1.5 eV in both bulk and ML CrI$_3$,
and thus this would give rise to the diminishing SOC matrix element of $\langle d_{z^2}|H_{SO}|d_{yz}\rangle^2$
which favors an in-plane magnetization.  
All these together would lead to the magneto-crystalline anisotropy energes in the CrI$_3$ structures 
(see Table 1) which strongly favor the out-of-plane magnetization. 

\subsection{Optical conductivity}
Now let us turn our attention to the calculated optical conductivity, which is the ingredient for evaluating 
the magneto-optical effects, as explained already in Sec. 2. The calculated optical conductivities of 
the investigated structures are displayed in Fig. 6 for bulk and ML CrI$_3$ as well as in Fig. S4 
in the SI for BL and TL CrI$_3$. Since the calculated optical conductivities for all the considered 
structures (see Figs. 6 and S4) are rather similar due to the weak van der Waals interlayer interaction, 
here we discuss only bulk and ML CrI$_3$ as examples. Notably, Fig. 6 shows that in both bulk and ML CrI$_3$,
the absorptive and dispersive parts of the diagonal optical conductivity elements for an 
in-plane electric polarization ($E || ab$) ($\sigma_{xx}$) differ significantly from that for 
the out-of-plane electric polarization ($E || c$) ($\sigma_{zz}$) above the absorption edge of about 1.5 eV. 
For example, the absorptive part $\sigma_{xx}^1$ for $E || ab$ of bulk CrI$_3$ is significantly larger than
that ($\sigma_{zz}^1$) for $E || c$ in the energy range from 1.5 to 4.4 eV, while it becomes smaller than
$\sigma_{zz}^1$ for $E || c$ above 4.4 eV. 
A similar profile is observed in ML CrI$_3$ in the energy region of 1.0 $\sim$ 4.3 eV, where 
higher value is found for $E || ab$ compared to that for $E || c$, while above 4.3 eV $\sigma_{zz}^1$ 
is higher than $\sigma_{xx}^1$. This manifests that these structures are highly optically anisotropic. 
The observed strong optical anisotropy is quite common in low dimensional materials and can be explained
in terms of the orbital-decomposed DOS spectra reported in the previous subsection. 
In particular, Figure 5(b) shows that the upper valance bands in the energy range of -4.5 to -0.3 eV are dominated 
by Cr $d_{xy,x^2-y^2}$ orbitals. Given that $d_{xy,x^2-y^2}$ ($d_{z^2}$) states can be excited by 
only E$\perp c$ (E$\parallel c$) polarized light while Cr $d_{xz,yz}$ orbitals can be excited by light 
of both polarizations, consequently $\sigma_{xx}^1 $ would be greater than $\sigma_{zz}^1$ in the 
low photon energy region up to 4.3 eV [see Figs. 6(a) and 6(b)]. 

The calculated real ($\sigma_{xy}^1$) and imaginary ($\sigma_{xy}^2$) parts of the off-diagonal optical 
conductivity element are displayed in Fig. 6(e) and Fig. 6(f) for bulk and ML CrI$_3$, respectively. 
It is quite evident from these figures that the off-diagonal element spectra of bulk and ML CrI$_3$ are again
rather similar, due to the weak van der Waals interlayer coupling in bulk CrI$_3$. 
For example, the $\sigma_{xy}^1$ shows a large positive peak at 4.2 eV for the bulk and at 4.4 eV for the ML. 
In addition, there is another prominant peak at $\sim$3.5 eV for bulk and ML structures. 
In the $\sigma_{xy}^{2}$ spectra, there is a large positive peak in the vicinity of 4.8 eV for both structures. 
There are also negative peaks in both $\sigma_{xy}^1$ and $\sigma_{xy}^2$ spectra.
For example, a negative peak occurs in the vicinity of 1.3 eV in the $\sigma_{xy}^1$ spectrum 
and at 3.3 eV in the $\sigma_{xy}^{2}$ spectrum for bulk and ML structures. 

\begin{figure}[htb]
\begin{center}
\includegraphics[width=8.5cm]{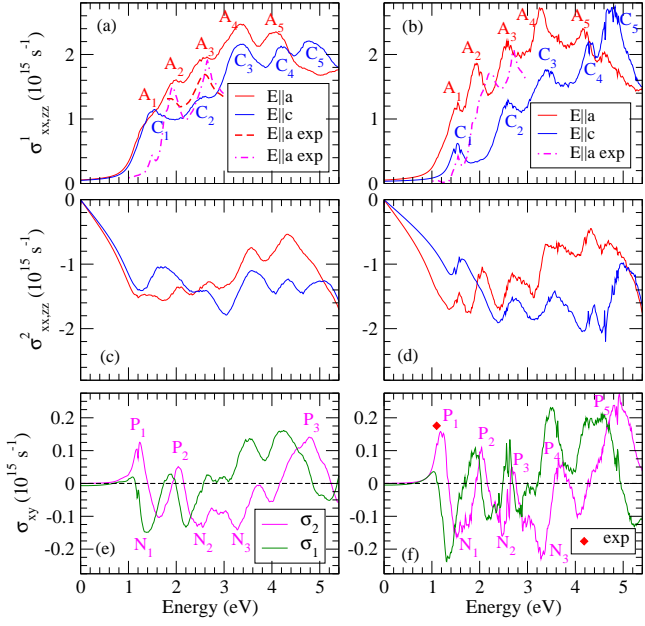}
\end{center}
\caption{(a) [(b)] Real, (c) [(d)] imaginary diagonal components and (e) [(f)] off-diagonal components 
of the optical conductivity tensor of bulk [ML] CrI$_3$. The pink dash-dotted lines in 
(a) and (b) are, respectively, the experimental differential reflection spectra 
(in abitrary units) for bulk and ML CrI$_3$~\cite{seyler-nat-2018}
which are proprotional to $\sigma_{xx}^1$. The red dashed curve in (a) is the $\sigma_{xx}^1$
of bulk CrI$_3$ derived from the experimental refraction index and extinction coefficent.~\cite{huang2017}
The red diamond in (f) denotes the $\sigma_{xy}^2$ of ML CrI$_3$ derived from the experimental
photoluminescence~\cite{seyler-nat-2018} (see text).
}	
\end{figure}

The absorptive parts of the optical conductivity elements ($\sigma_{xx}^1$, $\sigma_{zz}^1$, $\sigma_{xy}^2$ 
and $\sigma_{\pm}^1$) are directly related to the dipole allowed interband transitions, 
as  Eqs. (3), (4) and (10) indicate. Therefore, we further analyze the origin of the main features 
in the $\sigma_{xx}^1$, $\sigma_{zz}^1$ and $\sigma_{xy}^2$ spectra by considering the symmetries 
of the involved band states and hence the dipole selection rules (see Supplementary Note B in the SI). 
Since the spectra of $\sigma_{xx}^1$, $\sigma_{zz}^1$ and $\sigma_{xy}^2$ for all the considered systems 
are rather similar, as mentioned above, here we perform the symmetry analysis for ML CrI$_3$ only (see the SI). 
The scalar-relativistic and relativistic band structures of ML CrI$_3$ with the symmetries of 
band states at $\Gamma$-point labelled, are displayed in Fig. S6 and Fig. 7, respectively. 
We then assign the main peaks of the $\sigma_{xx}^1$ and $\sigma_{zz}^1$ spectra [Fig. 6(b)] 
to the interband transitions at the $\Gamma$-point using the dipole selection rules (see Table S3
in the SI), as presented in Fig. S6. For example, we could relate the A$_1$ peak 
at $\sim$1.5 eV in $\sigma_{xx}^1$ in Fig. 6(b) to the interband transition from the $\Gamma _3^+$ 
state at -0.5 eV of the spin-down valence band to the conduction band state $\Gamma _3^-$ at 1.2 eV. 
Of course, in addition to these dipole-allowed interband transitions at the $\Gamma$-point, 
there are also contributions from different interband transitions at other k-points. 
It should be noted that, without the SOC, the $\Gamma _3^+$ and $\Gamma _3^-$ band states 
are doubly degenerate (Fig. S6), and the absorption rates for left- and right-handed polarized lights 
are the same, thereby resulting in zero contribution to $\sigma_{xy}$. In the presence of the SOC, 
the band states split (see Fig. 7) and this gives rise to optical magnetic circular dichroism. 
For example, we could relate the peak P$_1$ at 1.2 eV in the $\sigma_{xy}^2$ of Fig. 6(f) to 
the interband transition from the $\Gamma _6^-$ occupied states at -0.1 eV  
to the $\Gamma _4^+$ unoccupied state at $\sim$1.1 eV (Fig. 7).

\begin{figure}[htb]
\begin{center}
\includegraphics[width=8cm]{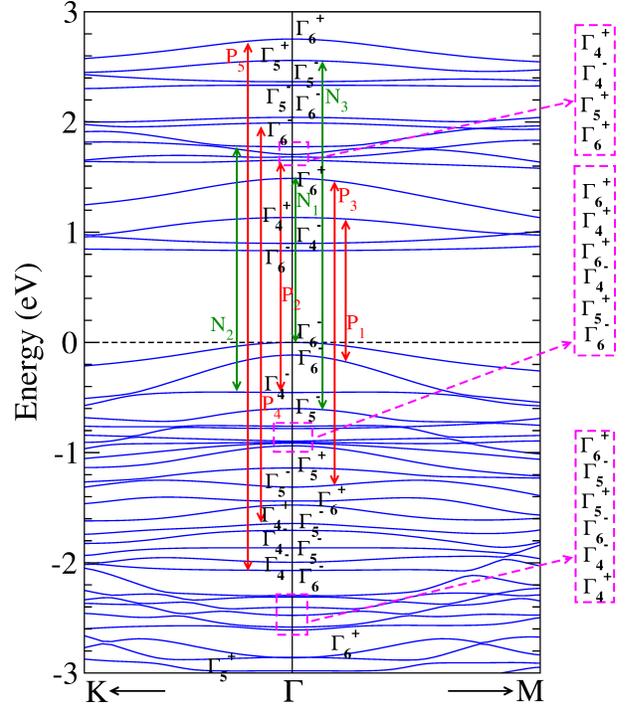}
\end{center}
\caption{Relativistic band structures of ML CrI$_3$. Horizontal dashed lines denote the top of valance band. 
The states at the $\Gamma$-point in the band structure are labeled according to the irreducible representation 
of the $C_{3i}$ double point group. The principal interband transitions and the corresponding peaks in 
the $\sigma_{xy}$ in Fig. 6 (f) are indicated by red and magenta arrows.}
\end{figure}

For comparison with the available experimental data, we plot in Figs. 6 (a) and (b) 
the experimental differential reflection spectra (in abitrary units) of bulk and ML CrI$_3$, 
respectively~\cite{seyler-nat-2018}, which are proportional to $\sigma_{xx}^1$. 
Moreover, we also display the $\sigma_{xx}^1$
of bulk CrI$_3$ derived from the experimental refraction index and extinction coefficent~\cite{huang2017}
in Fig. 6(a). It can be seen from Fig. 6(a) that twe peaks (A$_2$ and A$_3$ in 
the experimental $\sigma_{xx}^1$ of bulk CrI$_3$~\cite{huang2017} agree 
well with the theoretical ones in both shape and position, although the magnitude of
the experimental spectrum is slightly smaller. Moreover, three peaks (A$_1$, A$_2$ and A$_3$) in the 
experimental differential reflection spectrum of bulk CrI$_3$~\cite{seyler-nat-2018} 
also agree very well with the theoretical ones [see Fig. 6(a)]. 
This indicates that our GGA + $U$ calculations could describe the optical properties of 
bulk CrI$_3$ quite well. Figure 6(b) shows that three peaks (A$_1$, A$_2$ and A$_3$) in the 
experimental differential reflection spectrum of ML CrI$_3$~\cite{seyler-nat-2018}
are also quite well reproduced by our GGA + $U$ calculations. In particular, 
the position and shape of peak A$_1$ in theoretical and experimental spectra are
in good agreement, although theoretical peaks A$_2$ and A$_3$ 
are red-shifted by about 0.3 eV relative to the experimental ones.
Assuming photoluminescence (PL) intensity ($I$) is proportional to photoabsorption ($\sigma$),
PL circular polarization 
$\rho = (I_+-I_-)/(I_++I_-)\approx (\sigma_+^1-\sigma_-^1)/(\sigma_+^1+\sigma_-^1) = \sigma_{xy}^2/\sigma_{xx}^1$,
i.e., $\sigma_{xy}^2=\rho \sigma_{xx}^1$. In Fig. 6(f), the red diamond denotes 
the $\sigma_{xy}^2$ estimated this way based on the calculated $\sigma_{xx}^1$
and experimental $\rho$ for ML CrI$_3$~\cite{seyler-nat-2018}, which agree well with
peak P$_1$ in the theoretical $\sigma_{xy}^2$ spectrum.

Table 1 indicates that the calculated band gap of bulk CrI$_3$ is about 50 \% smaller
than the experimental value~\cite{dillon1965}. As usual, we could attribute 
this significant discrepancy to the known underestimation of the band gaps by the GGA
functional  where many-body effects especially quasiparticle self-energy correction are neglected. 
Since Eqs. (3) and (4) indicate that correct band gaps would be important
for obtaining accurate optical conductivity and hence magneto-optical effects,
we further perform the band structure calculations using the hybrid
Heyd-Scuseria-Ernzerhof (HSE) functional~\cite{heyd2003j,heyd2006} which is known
to produce improved band gaps for semiconductors (see Supplementary Note C in the SI for the details). 
The HSE band structures and band gaps are presented in Fig. S7 and Table S4 in the SI, respectively.
Indeed, the HSE band gap of bulk CrI$_3$ agrees well with the experimental value~\cite{dillon1965}.
Therefore, we also calculate the optical and magneto-optical properties of bulk and few-layer CrI$_3$
from the GGA + $U$ band structures within the scissors correction (SC) scheme~\cite{levine_prb_1991}
using the differences between the HSE and GGA band gaps (Table S4).
The optical conductivity spectra calculated with the SC are displayed in Fig. S8 in the SI. 
Surprisingly, the optical conductivity spectra of bulk and ML CrI$_3$
calculated with the SC are in worse agreement with the corresponding experimental
values~\cite{seyler-nat-2018,huang2017} [see Figs. S8(a), S8(b) and S8(j)].
This could be explained as follows.
We notice that bulk CrI$_3$ has an indirect band gap (see Fig. 3).
Consequently, the discrepancy in the band gap between theory and experiment in bulk CrI$_3$ could be attributed
to the fact that the experimental band gap is the optical one due to
direct interband transitions whose energy should be larger than the indirect band gap (see Fig. 7). 

\subsection{Magneto-optical Kerr and Faraday effects}
Finally, let us move onto the MO Kerr and Faraday rotational angles calculated
from the obtained optical conductivity spectra. To calculate the Kerr angles of the few-layer structures, 
we also need to know the dielectric constant of the substrate [see Eq. (8)].
Here we assume that the substrate is SiO$_2$ with dielectric constant $\varepsilon_s = 2.25$ 
for all CrI$_3$ multilayers since SiO$_2$ was used in the MOKE experiments~\cite{huang2017}. 
The calculated complex Kerr and Faraday rotation angles as a function of photon energy are presented,
respectively, in Figs. 8 and 9 for all the investigated structures. Figures 8 and 9 show that 
the MO spectra for all the investigated structures except the bulk, are negligible below the band gap 
of $\sim$1.0 eV. Nonetheless, they all become pronounced above 1.0 eV and oscillate as 
the photon energy increases. 
We notice that the Kerr rotation ($\theta_K$) and Kerr ellipticity ($\varepsilon_K$) 
of all the structures (Fig. 8) are, respectively, similar to that of the real part ($\sigma_{xy}^1$) 
and imaginary part ($\sigma_{xy}^2$) of the off-diagonal conductivity element (Fig. 6). 
This could be expected because the Kerr effect and the off-diagonal conductivity element 
are connected through Eqs. (7) and (8). In fact Eqs. 7 and 8 indicate that the complex Kerr rotation 
angle would be proportional to the off-diagonal optical conductivity of $\sigma_{xy}$ 
if the conductivity $\sigma_{xx}$ of the sample and substrate is roughly a  
constant of photon energy. For the CrI$_3$ multilayers, the dielectric constant of $\varepsilon = 2.25$ 
of the SiO$_2$ substrate is a constant and thus the Kerr rotation
angle is proportional to the $\sigma_{xy}$. 
For bulk CrI$_3$, the Kerr rotation could become large if the $\sigma_{xx}$, which is 
in the denominator of Eq. (7), becomes very small. 
This explains why the Kerr angles of the bulk are still pronounced below 0.5 eV [Fig. 8(a)].

\begin{figure}[htb]
\begin{center}
\includegraphics[width=7.5cm]{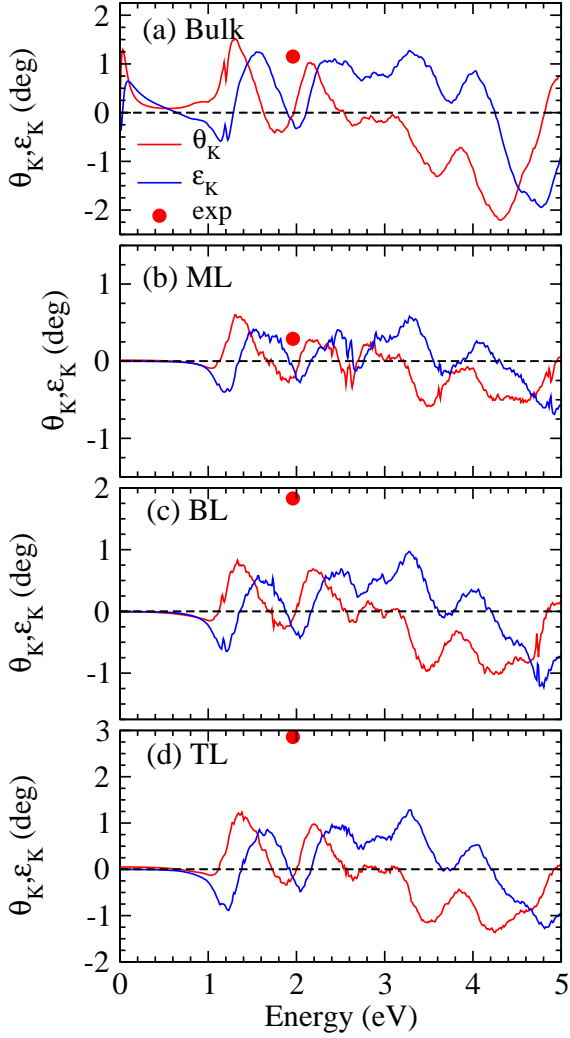}
\end{center}
\caption{Kerr rotation ($\theta_K$) and ellipticity ($\varepsilon_K$) spectra of (a) bulk, (b) ML, (c) BL,
and (d) TL CrI$_3$. The filled red circle in each panel indicates the $\theta_K$ value from 
the recent experiment~\cite{huang2017}}	
\end{figure}

The overall behavior of the MO spectra for all the investigated structures (Figs. 8 and 9) 
is rather similar, which could be attributed to the weakness of the van der Waals interlayer interaction. 
Notably, the maximum Kerr rotation angles for all the systems are large. Specifically, the maximum $\theta_K$
values for bulk, ML, BL and TL CrI$_3$ are, respectively, 1.5${\degree}$, 0.6${\degree}$, 0.8${\degree}$ 
and 1.2${\degree}$ at $\sim$1.3 eV. 
Kerr ellipticity shows a maximum value of 1.3${\degree}$ at 1.6 eV for the bulk as well as 
0.6${\degree}$, 0.9${\degree}$ and 1.2${\degree}$ at 3.3 eV for the ML, BL and TL, respectively. 
As for the calculated optical conductivities, we notice that the negative peaks also exist 
in both Kerr rotation and ellipticity spectra. The negative maximum of the Kerr rotation angle 
occurs at 4.3 eV for the bulk (-2.1${\degree}$) and at 4.2 eV for the TL (-1.3${\degree}$) as well as
at 3.3 eV for the ML (-0.5${\degree}$) and 4.2 eV for the BL (-1.0${\degree}$).
Similarly, Kerr ellipticity shows the negative maximum near 4.8 eV for the bulk (-1.9${\degree}$)
as well as at 4.9 eV for the ML (-0.6${\degree}$) and at 4.8 eV for BL (-1.2${\degree}$) and TL (-1.3${\degree}$).
These large Kerr angles and ellipticities indicate that the considered structures would have
valuable applications in magneto-optical nanodevices.

To access the application potential of these CrI$_3$ materials for applications, let us now
compare their MO properties with several well-known MO materials such
as 3$d$ transition metal alloys and compound semiconductors.
Among the transition metal alloys, manganese-based pnictides are known to have remarkable MO properties.
In particular, MnBi thin films were reported to have a large Kerr rotation angle of 2.3${\degree}$ 
at 1.84 eV.~\cite{ravindran_1999,di1996optical}  
Pt alloys also possess large Kerr rotations in the range of 0.4-0.5${\degree}$ in 
FePt and Co$_2$Pt ~\cite{gyg_prb51_1995} and PtMnSb~\cite{van1983ptmnsb}. 
In these materials, the strong SOC of heavy Pt atoms was shown to play an important 
role~\cite{guo_jmmm_1996}. For comparison, we note that 3$d$ transition metal multilayers 
were found to have rather small Kerr rotation angles of 0.06${\degree}$ at 2.2 eV 
for Fe/Cu multilayer, of 0.16${\degree}$ at 2.5 eV for Fe/Au 
multilayer~\cite{katayama_1988,suzuki_1992} and also of 0.06${\degree}$ 
at 1.96 eV for Co/Au multilayer~\cite{megy_1995}. 
Among semiconductor MO materials, diluted magnetic semiconductors Ga$_{1-x}$Mn$_x$As 
were reported to show significant Kerr rotation angles of $\sim$0.4${\degree}$ near 
of 1.80 eV~\cite{lang2005}. Y$_3$Fe$_5$O$_{12}$ also exhibits a significant Kerr rotation 
of 0.23${\degree}$ at 2.95 eV~\cite{tomita2006}. 
In strong contrast, the measured Kerr rotation angles of bulk and atomically thin films of Cr$_2$Ge$_2$Te$_6$
are much smaller~\cite{gong2017}. For example, the experimental Kerr angle at 0.8 eV is only $\sim$0.14${\degree}$ 
for the bulk, and ranges from 0.0007${\degree}$ in bilayer to 0.002${\degree}$ 
in trilayer~\cite{gong2017}. 
Overall, the Kerr rotation angles predicted for bulk and multilayer CrI$_3$ here are 
comparable to these important MO metals and semiconductors. Thus, because of their excellent MO properties, 
CrI$_3$ structures could have promising applications in, e.g., MO nanosensors and high density MO data-storage devices. 

\begin{figure}[htb]
\begin{center}
\includegraphics[width=7.5cm]{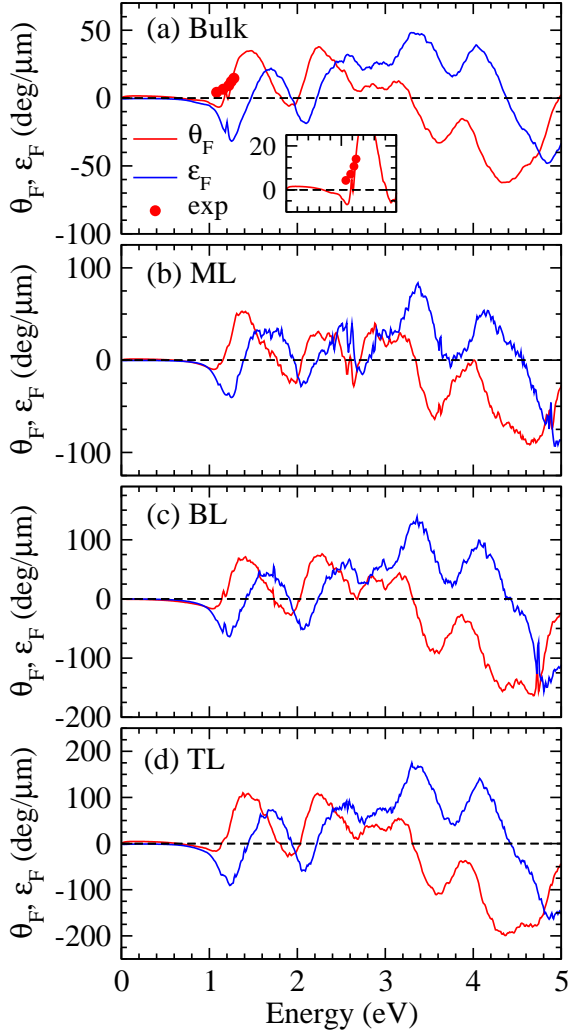}
\end{center}
\caption{Faraday rotation ($\theta_F$) amd ellipticity ($\varepsilon_F$) spectra of (a) bulk, (b) ML, (c) BL, 
and (d) TL CrI$_3$. The filled red circles in (a) denote the measured $\theta_F$ values of bulk CrI$_3$~\cite{dillon1966}.}	
\end{figure}

The calculated complex Faraday rotation angles for all the investigated structures are presented 
in Fig. 9. The main features of the Faraday rotation spectra of the considered structures resemble 
that of the Kerr rotation spectra (see Fig. 8). Notably, the large Faraday rotation maximum occurs  
at the photon energy of $\sim$2.3 eV for the bulk (36{\degree/$\mu$m}), of $\sim$1.4 eV 
for the ML (50{\degree/$\mu$m}), of 2.3 eV for the  BL (75{\degree/$\mu$m}) and TL (108{\degree/$\mu$m}).
Faraday ellipticity shows the large maximum near 3.3 eV for the bulk (48{\degree/$\mu$m}), 
the ML (80{\degree/$\mu$m}),  the BL (130{\degree/$\mu$m}) and the TL (170{\degree/$\mu$m}).  
Also we notice that the large negative maximum exists in both Faraday rotation and ellipticity spectra. 
For example, the negative maximum of the Faraday rotation 
occurs at 4.3 eV for the bulk (60{\degree/$\mu$m}) and the TL (-194{\degree/$\mu$m})
as well as at 4.7 eV for the ML (-85{\degree/$\mu$m}) and the BL (-155{\degree/$\mu$m}).
Faraday ellipticity shows the large negative maximum near 4.8 eV for all the structures 
with $\varepsilon_F$ = -47{\degree/$\mu$m}, -85{\degree/$\mu$m}, -148{\degree/$\mu$m} 
and -155{\degree/$\mu$m} for the bulk, ML, BL and TL, respectively. 
For comparison, we notice that the MnBi films are known to possess the largest Faraday rotation 
angles of $\sim$80\degree/$\mu$m at 1.77 eV~\cite{di1996optical,ravindran_1999}. 
Ferromagnetic semiconductor Y$_3$Fe$_5$O$_{12}$ show only small Faraday rotation angles
of $\sim 0.19${\degree/$\mu$m} at 2.07 eV~\cite{boudiar2004}. However, the replacement of Y with Bi 
considerably enhances the Faraday rotations to $\sim$35.0{\degree/$\mu$m} at 2.76 eV~\cite{vertruyen2008curie},
thus making Bi$_3$Fe$_5$O$_{12}$ an outstanding MO material. 
Clearly, the calculated 
Faraday angles of the CrI$_3$ multilayers are high compared to that of prominent bulk 
MO metals such as manganese pnictides. 

The Kerr rotation angles $\theta_K$ of bulk and atomically thin film CrI$_3$ have recently been
measured at photon energy of 1.96 eV \cite{huang2017} and they are found to be large (see Fig. 8).
If the calculated Kerr rotation spectra are red-shifted by merely 0.2 eV, 
the measured and calculated $\theta_K$ values of bulk and ML CrI$_3$ would be in very good agreement,
although the experimental $\theta_K$ values of BL and TL CrI$_3$ are much larger than
the corresponding theoretical $\theta_K$ values (Fig. 8). 
Furthermore, the measured Faraday rotation angles $\theta_F$ of bulk CrI$_3$\cite{dillon1966} 
are almost in perfect agreement with our theoretical prediction [see Fig. 9(a)].
We notice that in contrast, the photon energies of the experimental Faraday angles of bulk CrI$_3$\cite{dillon1966} 
differ significantly from that of the theoretical Faraday angles calculated with the scissors correction
using the HSE band gap [see Fig. S10(a) in the SI].

\section{Conclusion}
Summarizing, we have systematically studied magnetism, electronic structure, optical 
and magneto-optical properties of bulk and atomically thin films of CrI$_3$ 
by performing extensive GGA + $U$ calculations. 
Firstly, we find that both intralayer and interlayer magnetic couplings are strongly ferromagnetic
in all the considered CrI$_3$ structures. Ferromagnetic ordering temperatures estimated 
based on calculated exchange coupling parameters agree quite well with the measured ones.
Secondly, we reveal that all the structures have a large out-of-plane MAE of $\sim$0.5 meV/Cr,
in contrast to the significantly thickness-dependent MAE in atomically thin films of Cr$_2$Ge$_2$Te$_6$.~\cite{gyg_prb2018}
These large MAEs suppress transverse spin fluctuations in atomically thin CrI$_3$ films
and thus stabilize long-range magnetic orders at finite temperatures down to the ML limit.
Thirdly, all the structures also exhibit strong MO Kerr and Faraday effects.
In particular, in these CrI$_3$ structures, MO Kerr rotations up to $\sim$2.6$\degree$ are found,
being comparable to famous MO materials such as ferromagnetic semiconductor Y$_3$Fe$_5$O$_{12}$ and metal PbMnSb.
Calculated MO Faraday rotation angles are also large, being up to $\sim$200{\degree/$\mu$m},
and they are comparable to that of best known
MO semiconductors such as Y$_3$Fe$_5$O$_{12}$ and Bi$_3$Fe$_5$O$_{12}$.
Calculated optical conductivity spectra, MO Kerr and Faraday rotation angles agree quite well
with available experimental data. 
In BL CrI$_3$, the interlayer magnetic coupling is found to be layer stacking dependent
while the optical and magneto-optical properties are not.
Therefore, further experimental determination of the structure of BL CrI$_3$ is needed to resolve
why the antiferromagnetic state instead of the predicted ferromagnetic state, is observed.
Fourthly, all the structures are found to be single-spin ferromagnetic semiconductors.
Finally, calculated MAEs, optical and magneto-optical conductivity
spectra of these structures are analyzed in terms of their underlying electronic band structures.
Our findings of large out-of-plane MAEs and strong MO effects in these single-spin ferromagnetic
semiconducting CrI$_3$ ultrathin films suggest that they will find promising applications in high density
semiconductor MO and low-power spintronic nanodevices as well as efficient spin-photovoltaics.

\textbf{Acknowledgement:}
V. K. G. and G. Y. G. acknowledges the support by the Ministry of Science and Technology 
and the National Center for Theoretical Sciences, Taiwan. G. Y. G. also thanks 
the Academia Sinica and the Kenda Foundation of Taiwan for the supports.

\textbf{References:}
\label{refs}
{}

\end{document}